\documentclass[preprint,12pt]{elsarticle}

\usepackage{graphicx}
\usepackage{amssymb}

\usepackage{lineno}

\usepackage{url}

\journal{Physica Medica, European Journal of Medical Physics}

\begin{document}

\begin{frontmatter}


\title{Simulation of a radiobiology facility for the Centre for the Clinical Application of Particles}



\author[IC]{A. Kurup\corref{cor1}}
\ead{a.kurup@imperial.ac.uk}
\author[IC]{J. Pasternak}
\author[IC]{R. Taylor\fnref{Birmingham}}
\author[IC]{L. Murgatroyd\fnref{Birmingham}}
\author[JAI]{O. Ettlinger}
\author[RHUL]{W. Shields}
\author[RHUL]{L. Nevay}
\author[MUW]{S. Gruber}
\author[IC]{J. Pozimski}
\author[IC]{H. T. Lau}
\author[IC]{K. Long}
\author[IC]{V. Blackmore}
\author[IC]{G. Barber}
\author[JAI]{Z. Najmudin}
\author[ICR]{J. Yarnold}
\cortext[cor1]{Corresponding author.}

\address[IC]{Imperial College London, Exhibition Road, London SW7 2AZ, United Kingdom}
\address[JAI]{The John Adams Institute for Accelerator Science, Department of Physics, Blackett Laboratory, Imperial College London, London SW7 2AZ, United Kingdom}
\address[RHUL]{Royal Holloway, University of London,Egham Hill, Egham, Surrey TW20 0EX, United Kingdom}
\address[MUW]{Medical University of Vienna,Spitalgasse 23, 1090 Vienna, Austria}
\address[ICR]{The Institute of Cancer Research, 123 Old Brompton Road, London SW7 3RP, United Kingdom}
\fntext[Birmingham]{Visiting from University of Birmingham, Edgbaston, Birmingham B15 2TT, United Kingdom}

\begin{abstract}
The Centre for the Clinical Application of Particles' Laser-hybrid Accelerator for Radiobiological Applications (LhARA) facility is being studied and requires simulation of novel accelerator components (such as the Gabor lens capture system), detector simulation and simulation of the ion beam interaction with cells. The first stage of LhARA will provide protons up to 15\,MeV for in vitro studies. The second stage of LhARA will use a fixed-field accelerator to increase the energy of the particles to allow in vivo studies with protons and in vitro studies with heavier ions.

BDSIM, a Geant4 based accelerator simulation tool, has been used to perform particle tracking simulations to verify the beam optics design done by BeamOptics and these show good agreement. Design parameters were defined based on an EPOCH simulation of the laser source and a series of mono-energetic input beams were generated from this by BDSIM. The tracking results show the large angular spread of the input beam (0.2\,rad) can be transported with a transmission of almost 100\% whilst keeping divergence at the end station very low ($<0.1$\,mrad).  The legacy of LhARA will be the demonstration of technologies that could drive a step-change in the provision of proton and light ion therapy (i.e. a laser source coupled to a Gabor lens capture and a fixed-field accelerator), and a system capable of delivering a comprehensive set of experimental data that can be used to enhance the clinical application of proton and light ion therapy.
\end{abstract}


\end{frontmatter}


\section{Introduction}
\label{sec:intro}


Cancer is a major cause of death with 17 million new cases each year globally and incidence rates predicted to increase to 27.5 million new cases per year by the year 2040~\cite{cancer_research_uk_2018}.  Radiotherapy is an important treatment modality and in England, 27\% of patients received radiotherapy as part of their primary cancer treatment during 2013--2014~\cite{cancer_research_uk_2017}.  Conventional radiotherapy uses X-ray photons to deliver dose to the tumour, causing damage that can kill the cancerous cells.  The dose deposition distribution of X-rays as a function of depth within the patient is exponential. This means that treating tumours that are deep within the patient will deliver a higher dose to the organs and sensitive tissues located in front of the tumour than the dose delivered to the tumour.  This can be mitigated by using a number of different treatment fields, i.e. several treatments from different angles, to reduce the dose given to organs surrounding the tumour.  However, this is an important issue for tumours that are close to critical structures such as the brain and spinal cord and for infants where dose to healthy tissue, organs and bone can lead to developmental issues and a higher probability of causing secondary malignancies later in life. Protons and other light ion beams are attractive options for radiotherapy as, unlike photons, there is little (ions) or no (protons) dose deposited beyond the distal tumour edge.  In addition, as the dose is primarily deposited in the Bragg peak region near the maximum range of the beam for a given energy, treatment can be conformed to the tumour volume by using beams of different energies, which is referred to as a spread out Bragg peak (SOBP). The effect of using an SOBP is to increase the dose to tissue and organs in front of the tumour but sensitive organs behind the tumour would still only receive very little dose compared to treatment with photons. For an SOBP, the dose given to tissue in front of the tumour depends on the number of different beam energies needed to deliver the required dose to the tumour and on the depth of the tumour.  By carefully choosing the treatment fields, dose to sensitive organs can be reduced compared to an equivalent treatment with photons, thus improving patient outcomes.  See Figure 1 in~\cite{Levin2005} for an example of the dose depth profile of an SOBP and a photon dose profile for comparison.

The importance of proton beam therapy has been recognised by NHS, UK, with the creation of two new proton beam centres to treat teenagers and young adults with cancer and other patients with certain cancers such as highly complex brain, head and neck cancers~\cite{nhs_choices}.
However, the conventional methods used to produce proton beams at energies suitable for clinical use are expensive and require relatively large accelerators and a beam delivery system that requires precisely positioning large magnets in an arc such that the beam can be rotated around the patient.  In comparison, X-ray machines are compact and relatively cheap allowing them to be installed in a hospital room instead of requiring a dedicated facility.  It has been proposed by~\cite{BULANOV2002240},~\cite{Fourkal_2003}~and~\cite{Malka_2004} that laser-driven ion beams could provide an alternative to conventional accelerator facilities for radiotherapy applications.  Laser-driven ion beams can provide very high dose rates (from several hundred~Gy/s to more than $10^9$~Gy/s) and can produce carbon ions in addition to protons. Studies have indicated that radiotherapy with ultra-high dose rates can result in improved normal tissue sparing for the same total treatment dose, see~\cite{Favaudon245ra93}~and~\cite{Vozenin_2019}. Thus, laser-driven ion beams provide a very promising alternative to conventional accelerators. However, issues such as the maximum energy achievable, large beam divergence, large energy spread and shot-to-shot variations need to be addressed.

Another important issue for proton beam therapy is that treatment planning assumes that the relative biological effectiveness (RBE) is a factor of 1.1~\cite{Paganetti:2014}, which means a lower dose of protons is needed to produce the same effect than if photons were used.  However, this is an average value that can vary with several physical and biological parameters including particle type, dose, dose rate, linear energy transfer and biological endpoint. A number of other studies have also shown there can be significant variation in RBE, see~\cite{Jones:2018},~\cite{Giovannini2016}~and~\cite{Armin_2018}.  A detailed systematic study of the biophysical effects of the interaction of protons, under different physical conditions, with different tissue types would provide important information on RBE variation and could enable enhanced treatment planning.  Such studies are especially needed in the case of heavier ion beam radiotherapy. A number of studies done in the past have used laser-driven ion beams to investigate radiobiological effects, see for example~\cite{Kraft_2010},~\cite{Fiorini_2011},~\cite{Doria_2012},~\cite{Zeil_2013},~\cite{Masood_2014}~and~\cite{Zlobinskaya_2014}.  More recent projects (e.g. A-SAIL, ELI and SCAPA) will also study radiobiological effects using laser-driven ion beams and will address various technological issues, see for example~\cite{Manti_2017},~\cite{ROMANO2016153},~\cite{Masood_2017},~\cite{Chaudhary_2017}~and~\cite{Margarone_2018}.

The Centre for the Clinical Application of Particles (CCAP) at Imperial College London is composed of clinical oncologists, medical physicists, accelerator and instrumentation scientists and radiobiologists.  One focus of the CCAP's programme is the Laser-hybrid Accelerator for Radiobiological Applications (LhARA).  LhARA  will prove the principal of certain novel technologies for future therapy facilities by developing a proton and light ion beam radiobiology facility for in vitro and in vivo studies, which will contribute to the vibrant field of laser-driven ion beam radiobiology.  Conventional proton and ion sources produce particles with energies of tens of keV/u.  At such low energies, the Coulomb repulsion between the particles that make up the beam limit the beam-current that can be captured and accelerated.  Laser accelerated ions can be generated at higher energies ($1\times10^{4}$\,keV or more) and therefore do not suffer the same limitations, allowing injection into a strong focusing particle capture system.  One issue with the laser-driven ion beam is that it has a significant divergence making it very difficult to capture using conventional focusing elements. LhARA will use a Gabor lens, which is a confined-plasma focusing device, to improve the capture of the laser-driven ion beam. The beam can then be manipulated using conventional magnets and injected into a fixed-field accelerator (FFA), which has a large acceptance, that will accelerate the particles to energies suitable for in vivo experiments. Using these novel technologies for LhARA will make it possible to deliver multiple ion species from a single source, while overcoming the beam divergence issues of laser-driven ion beams, to provide a facility that can produce intense beams (and thus ultra-high dose rates) of protons and ions from helium to carbon to deliver a definitive programme of radiobiology.  The legacy of this programme will be the demonstration of technologies that can drive a step-change in the provision of proton and light ion therapy (i.e. a laser source coupled to a Gabor lens capture and an FFA), and a system capable of delivering a comprehensive set of experimental data that can be used to enhance the clinical application of proton and light ion therapy. 

 This paper will present the design concept of the first stage of LhARA that will use the laser source coupled to a Gabor lens capture system and a conventional beam-line (i.e using normal conducting magnets) to deliver proton beams (up to 15\,MeV) suitable for in vitro studies only. The second stage of LhARA that will deliver protons for in vivo experiments and heavier ion beams for in vitro experiments requires injection into an FFA that will increase the momentum by a factor of three (i.e. up to a kinetic energy of 127\,MeV for protons and 33\,MeV/u for carbon 6+ ions).  Initial tracking simulations of the first stage of LhARA using BDSIM~\cite{Agapov:2009zz} and~\cite{BDSIM:2018}, which is an accelerator simulation tool based on Geant4~\cite{ALLISON2016186} (the version used in these simulations is 10.4.1), have been performed to verify the optics design of the beam-line (done using BeamOptics~\cite{nla.cat-vn1446631}) and to determine the energy deposition profile in the end station to confirm that the 15\,MeV laser cut-off energy is suitable.

\section{The Laser-Hybrid Accelerator for Radiobiological Applications (LhARA)}
\label{sec:LhARA}
\begin{figure}[ht]
\centering\includegraphics[width=0.9\linewidth]{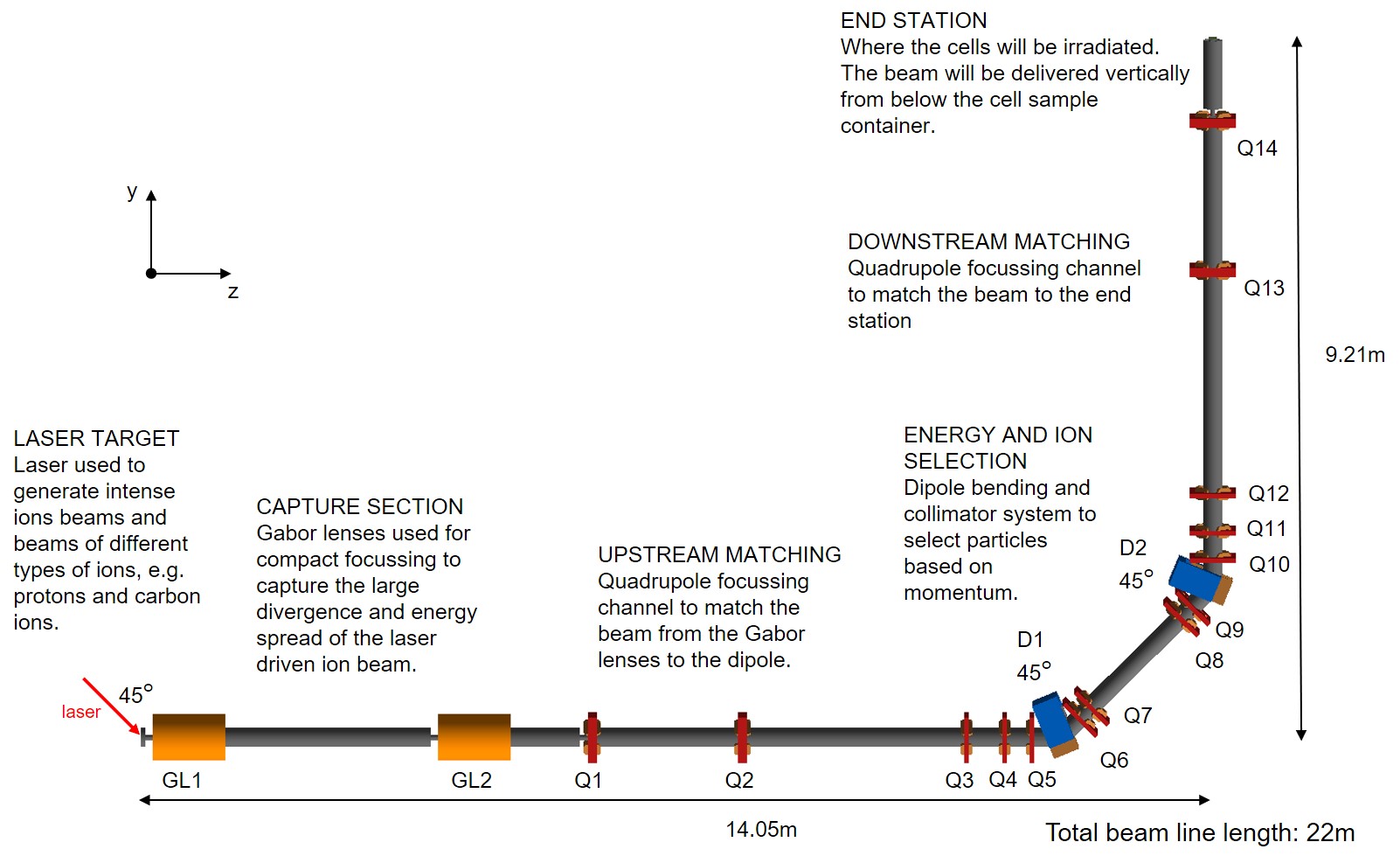}
\caption{The LhARA beam-line as implemented in BDSIM. The direction of the laser beam used to generate the beam is marked and the various sections are annotated.}
\label{fig:LhARA}
\end{figure}

The first stage of LhARA will use a laser-driven ion beam for detailed systematic studies of in vitro radiobiology.  Figure~\ref{fig:LhARA} shows the LhARA beam-line as implemented in BDSIM, which consists of: a capture system; upstream matching; a vertical 90$^{\circ}$ bend that provides energy and ion selection; downstream matching; and the end station which includes a vacuum window and a cell culture plate. The capture section is based on Gabor lenses that provide compact capture and focusing of the large divergence and large energy spread of the laser-driven ion beam. The upstream matching section is a quadrupole focusing channel used to match the beam from the capture section to the section that performs energy and ion selection. Two 45$^{\circ}$ dipole bends and collimators are used to select particles of the required momentum. The gap between the two dipoles is large enough to place a Wien filter for velocity selection, which would allow particle species selection in the case that separating particle species based on momentum is not effective enough.  The downstream matching section then transports the beam such that at the entrance of the end station (i.e. the vacuum window) the beam has a very small divergence ($<10^{-4}$\,rad), occupies the 30\,mm diameter aperture and has an energy spread of $\pm 2\%$.

\subsection{Laser source}
\label{sec:laser}
\begin{figure}[ht]
\centering\includegraphics[width=0.8\linewidth]{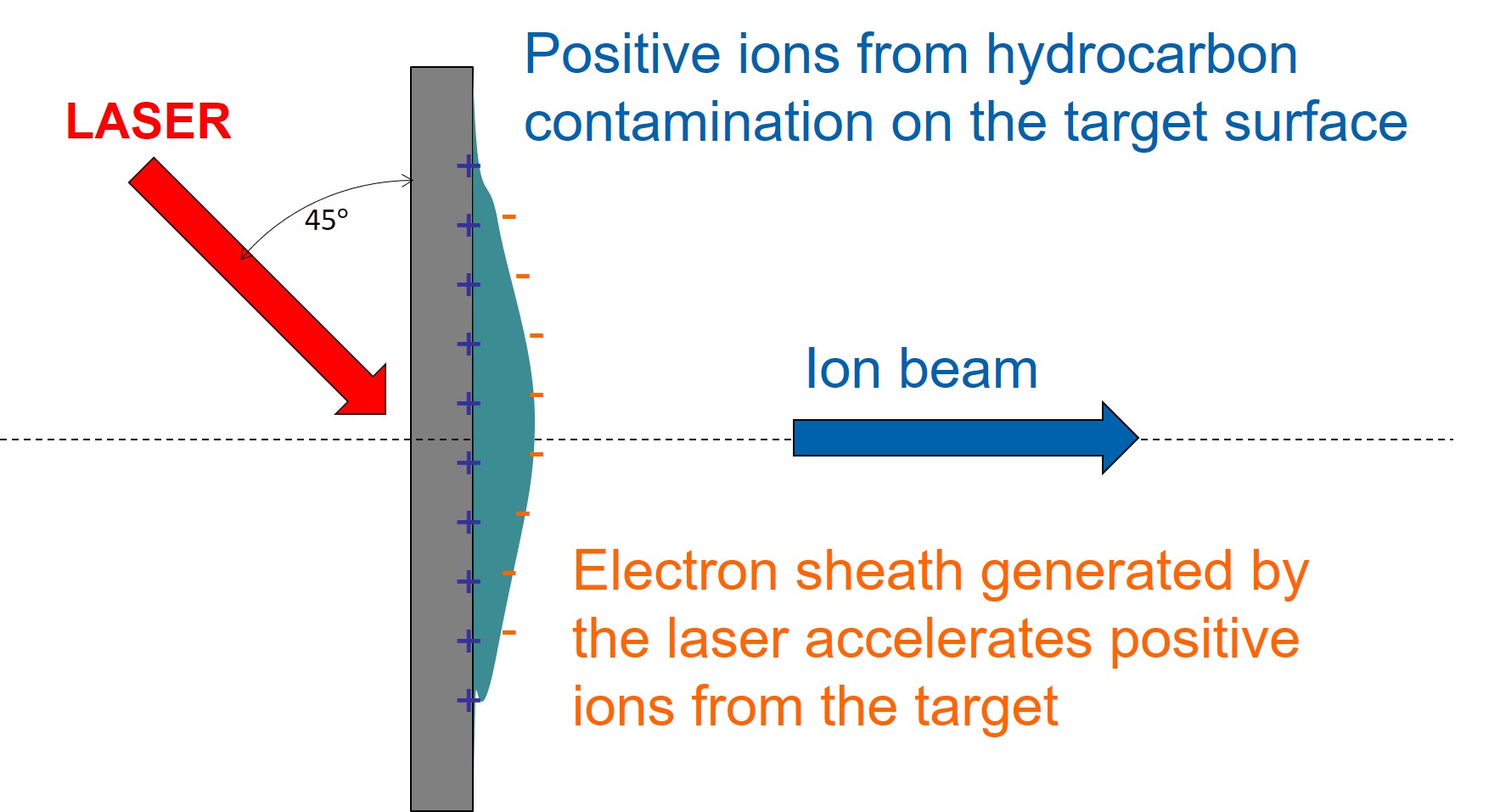}
\caption{Schematic diagram illustrating the principle used to generate the ion beam.}
\label{fig:laserPrinciple}
\end{figure}

\begin{figure}[ht]
\centering\includegraphics[width=0.8\linewidth]{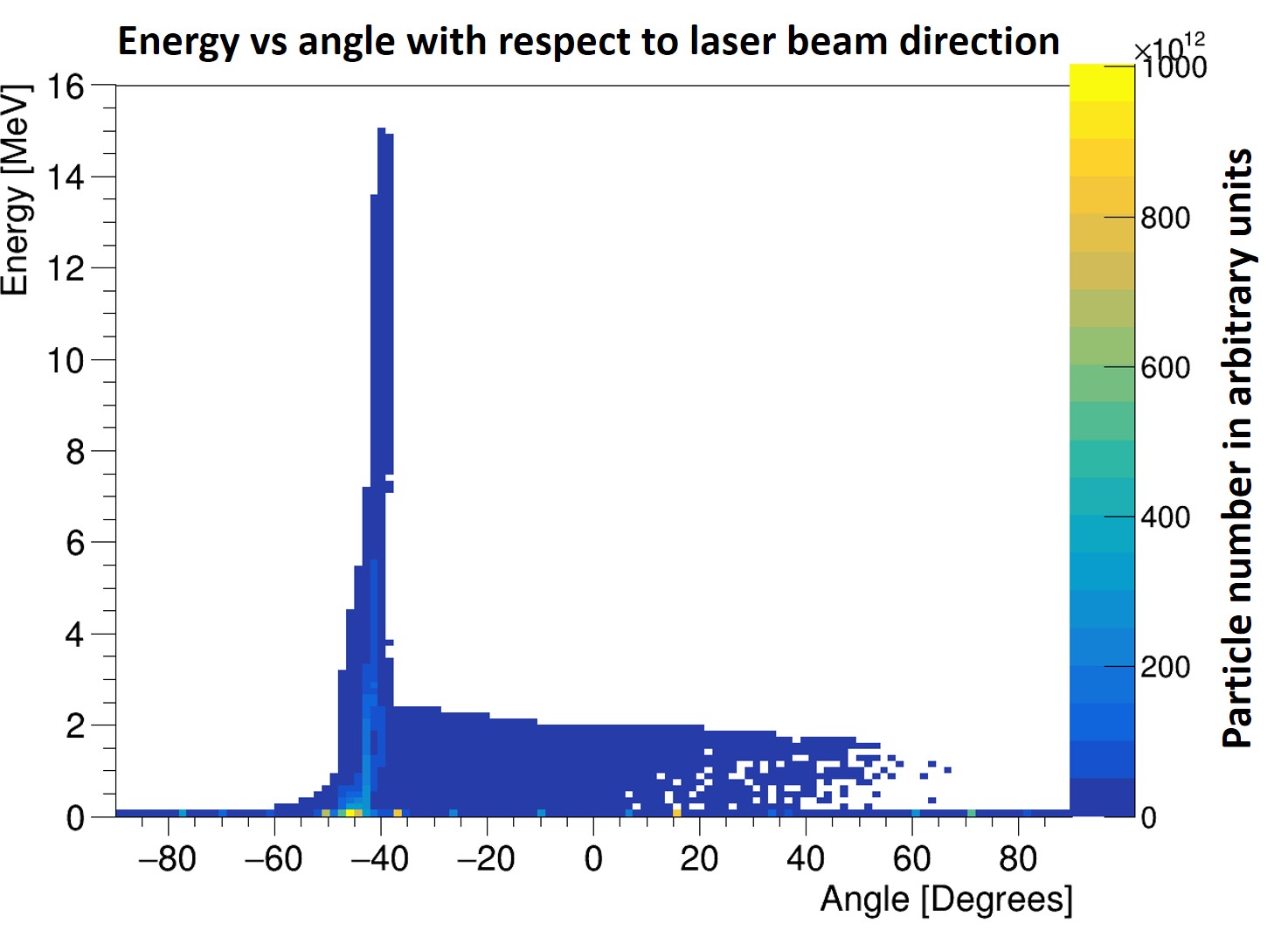}
\caption{Distribution of energy and angle from a proton beam simulation using EPOCH with a cut-off energy of 15\,MeV. In this simulation 0 degrees is the direction of the laser.}
\label{fig:laserBeam}
\end{figure}

The laser uses the principle of target normal sheath acceleration to generate an ion beam. Figure~\ref{fig:laserPrinciple} shows the basic principle where a high intensity laser pulse is fired at a thin target at an angle of 45$^{\circ}$ which causes an ion beam (consisting of protons and carbon ions) to be generated from the contaminants on the back surface of the target. See~\cite{Borghesi:2014} for a more detailed explanation of sheath acceleration. Using this method generates a high-intensity beam composed of a variety of ions but with a very large energy spread and large angular divergence. A typical energy and angle distribution of a proton beam with a cut-off energy of 15\,MeV, from an EPOCH~\cite{Arber:2015} simulation, is shown in Figure~\ref{fig:laserBeam}.

\begin{figure}[ht]
\centering\includegraphics[width=0.8\linewidth]{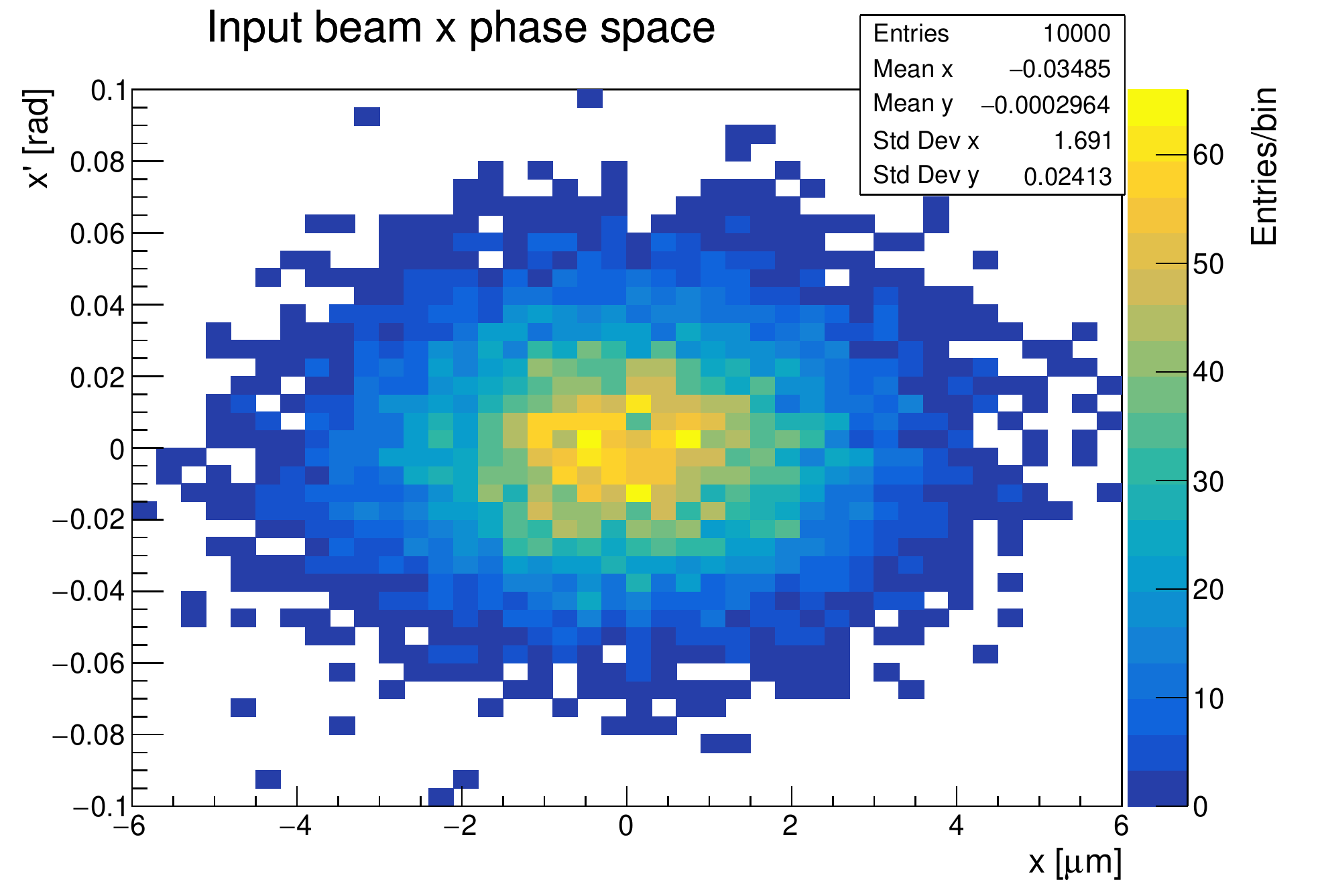}
\caption{Distribution of the x phase space of the input beam used in the BDSIM simulations.}
\label{fig:primXPhaseHist}
\end{figure}

\begin{figure}[ht]
\centering\includegraphics[width=0.8\linewidth]{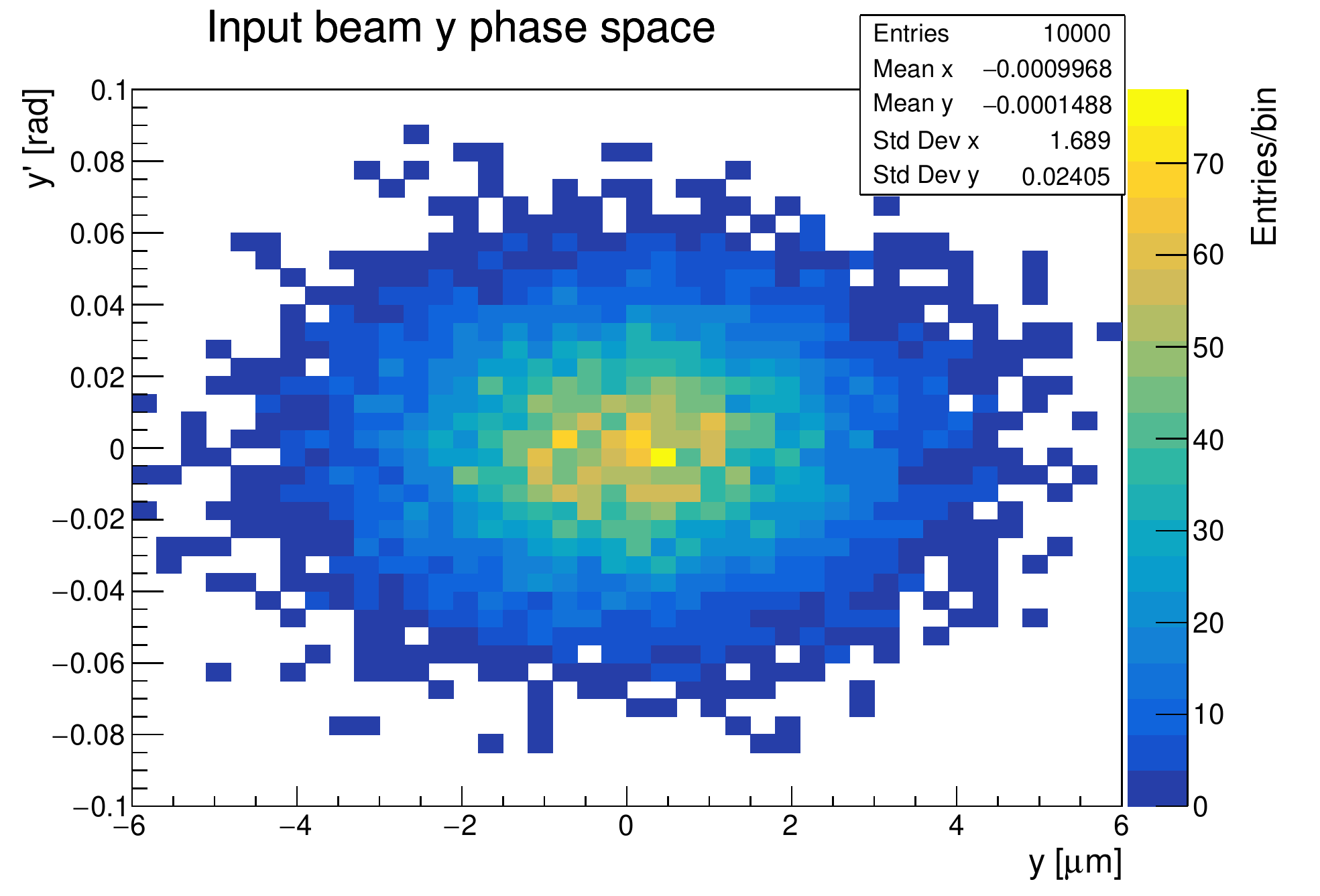}
\caption{Distribution of the y phase space of the input beam used in the BDSIM simulations.}
\label{fig:primYPhaseHist}
\end{figure}

\begin{figure}[ht]
\centering\includegraphics[width=0.8\linewidth]{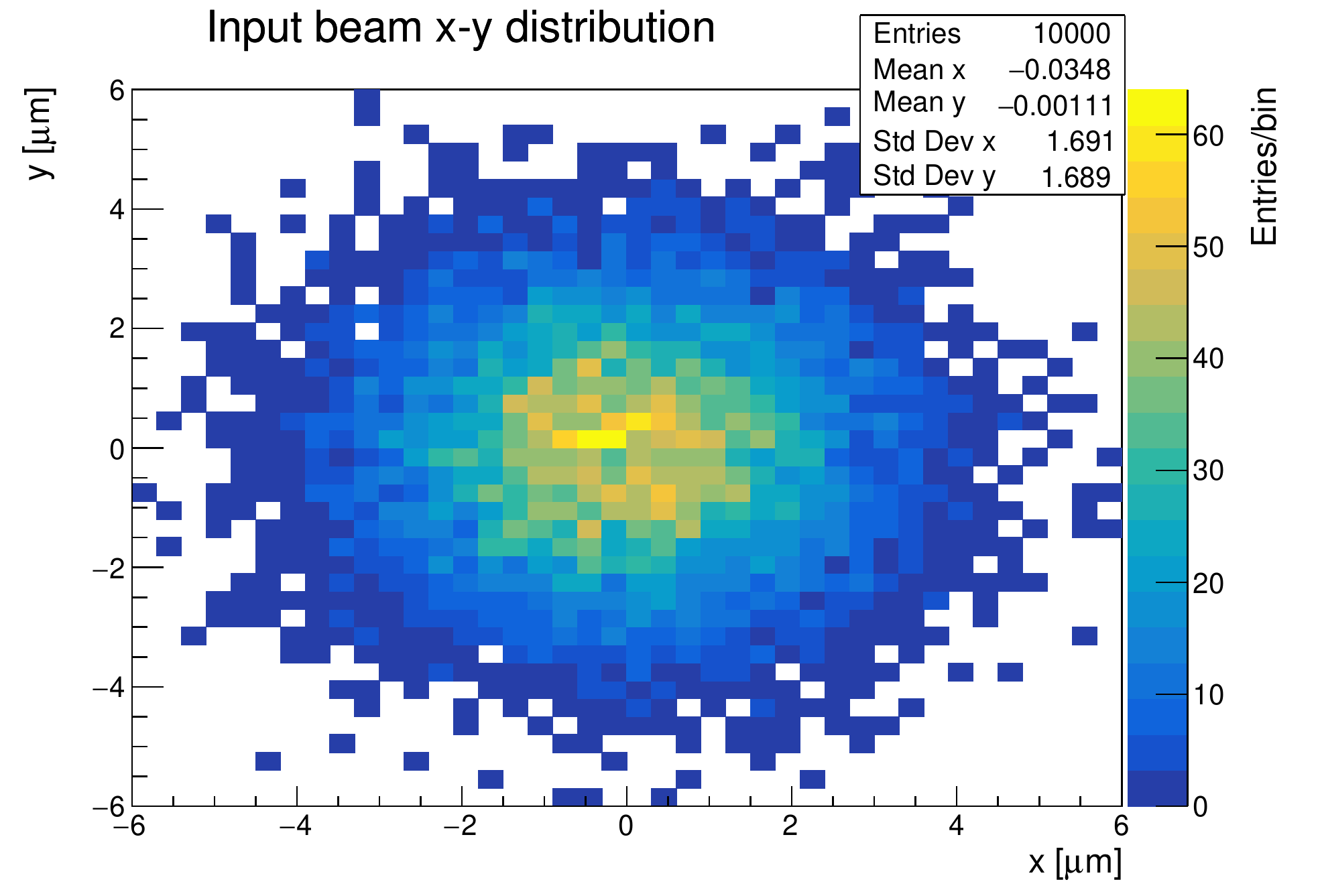}
\caption{Distribution of the x-y profile of the input beam used in the BDSIM simulations.}
\label{fig:primXYHist}
\end{figure}

The EPOCH simulation was used to determine the design parameters for the optics design, which are given in Table~\ref{tab:initialBeam}. These were used to generate a Gaussian distributed beam of $10^4$ protons using BDSIM in order to compare the tracking simulations with the optics design of the LhARA beam-line. The initial particle distribution used in the tracking simulations is shown in Figures~\ref{fig:primXPhaseHist},~\ref{fig:primYPhaseHist}~and~\ref{fig:primXYHist}.
The combination of a small Twiss beta and relatively large emittance means the beam has a small spot size but a very large divergence (the opening angle is 0.2\,rad).  
\begin{table}[ht]
\centering
\begin{tabular}{l l}
\hline
Alpha & 0 \\
Beta & $71\times 10^{-6}$\,m \\
Emittance & $4\times 10^{-8} \pi$\,m\,rad \\
Kinetic energy & 15\,MeV \\
\hline
\end{tabular}
\caption{Beam parameters used to generate an input beam for tracking with BDSIM. The parameters alpha and beta are the Twiss parameters.  Since the beam is assumed to be initially cylindrically symmetric, beta, alpha and the emittance are the same for both the x and y planes.}
\label{tab:initialBeam}
\end{table}


\subsection{Capture}
\label{sec:capture}
\begin{figure}[ht]
    \centering
    \includegraphics[width=0.9\linewidth]{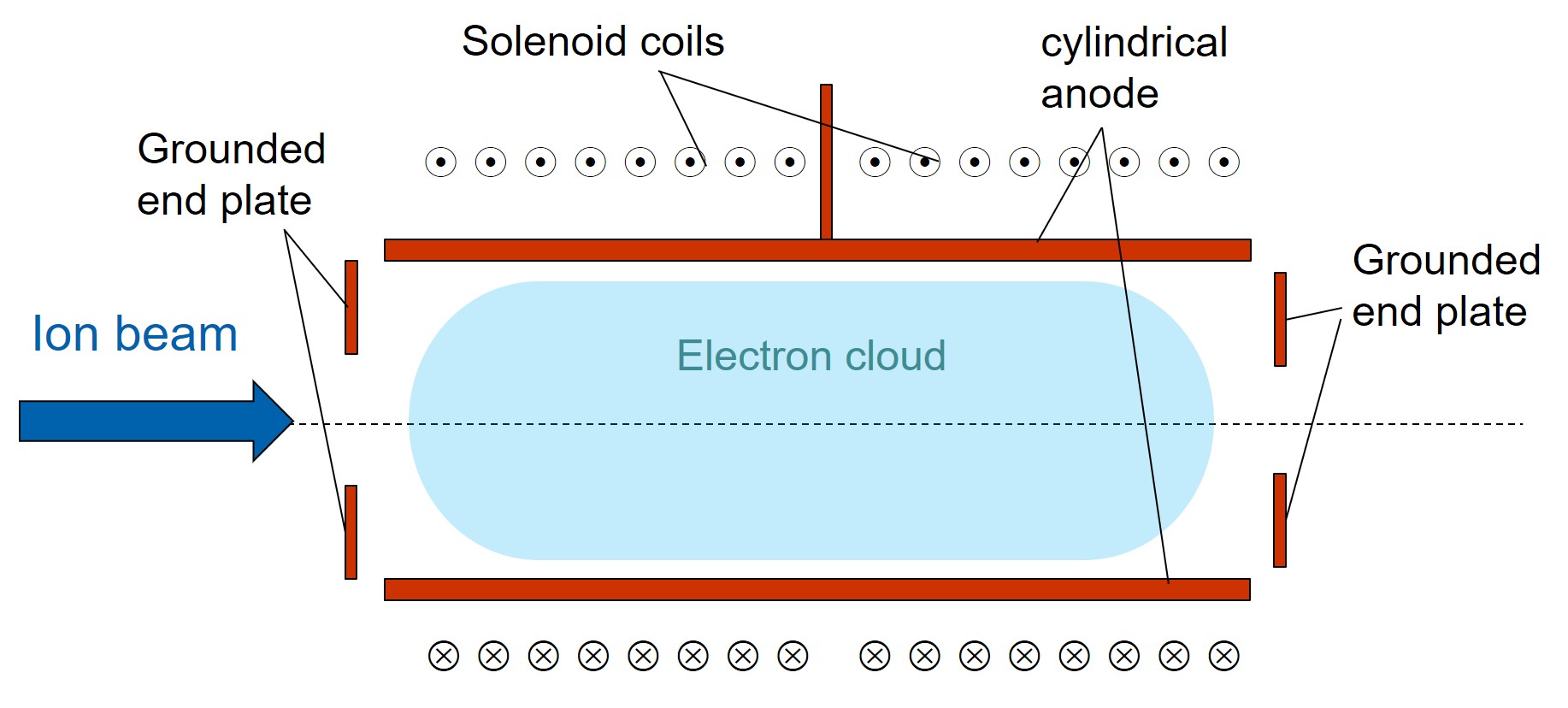}
    \caption{Schematic diagram of the Gabor lens.}
    \label{fig:gaborSchematic}
\end{figure}

The ion beam generated from the laser target is captured by a series of Gabor lenses.  A Gabor lens uses a confined electron plasma to produce an electro-static focusing field.  Figure~\ref{fig:gaborSchematic} shows the key components of the Gabor lens. The magnetic field produced by the solenoid coils causes the electrons to exhibit spiral trajectories parallel to the axis of the lens, thus confining them in the radial direction.  The voltage applied to the anode generates the space charge and the field between the anode and the ground plates at the ends confines the electron cloud in the longitudinal direction.

Details of the Gabor lens and its application to the capture of laser-driven ion beams for cancer therapy is described in~\cite{pozimski_aslaninejad_2013} and shows the expected opening angle of the laser source captured by the Gabor lens (for 200\,MeV protons and 5\% energy spread) to be around 76\,mrad which is better than a similar solenoid solution (requiring a 10\,T field), e.g. in~\cite{PhysRevSTAB.14.031304}, that would only capture 40\,mrad.  At these clinically relevant energies, the Gabor lens is expected to provide better performance without the need for strong magnetic fields and would be significantly cheaper. Thus, it is important that LhARA demonstrates the capability of using the Gabor lens to capture a laser-driven ion beam.

A prototype of the lens has been built and tested using a 1\,MeV proton beam at the Surrey Ion Beam Centre, see~\cite{Posocco:2016uea}.  Following these tests the prototype has been upgraded and is now being recommissioned at Imperial College London, UK.  It is planned that the upgraded lens will be tested using a laser-driven ion beam also at Imperial College London.

\subsection{Beam transport}
\label{sec:beamTransport}
The beam transport section includes the upstream matching, 90$^{\circ}$ vertical bend and the downstream matching, all of which use conventional quadrupole and dipole magnets.  The 90$^{\circ}$  bend is performed using two 45$^{\circ}$ dipole magnets where the drift between the magnets is long enough for collimators and an optional Wien filter to be inserted. 
The dispersion created by the dipoles can be used in conjunction with slit collimators to select only particles of a particular momentum, with the width of the slit determining the range of momenta selected.  The momentum spectrum of carbon ions produced by the target is expected to be lower than that of protons, allowing separation of protons from carbon ions purely based on momenta.  If however there is an overlap in the momentum spectrum of protons and other ion species there is sufficient space between the dipoles to include a Wien filter. This allows the selection of particles based on their velocity, by using electrostatic and magnetic deflectors and a collimator.  Work is ongoing to understand better the momentum spectrum of the ions produced by the target taking into account fragmentation effects to verify that selecting a pure sample of protons can be achieved.

\subsection{End station}
\label{sec:end}
\begin{figure}[ht]
    \centering
    \includegraphics[width=0.6\linewidth]{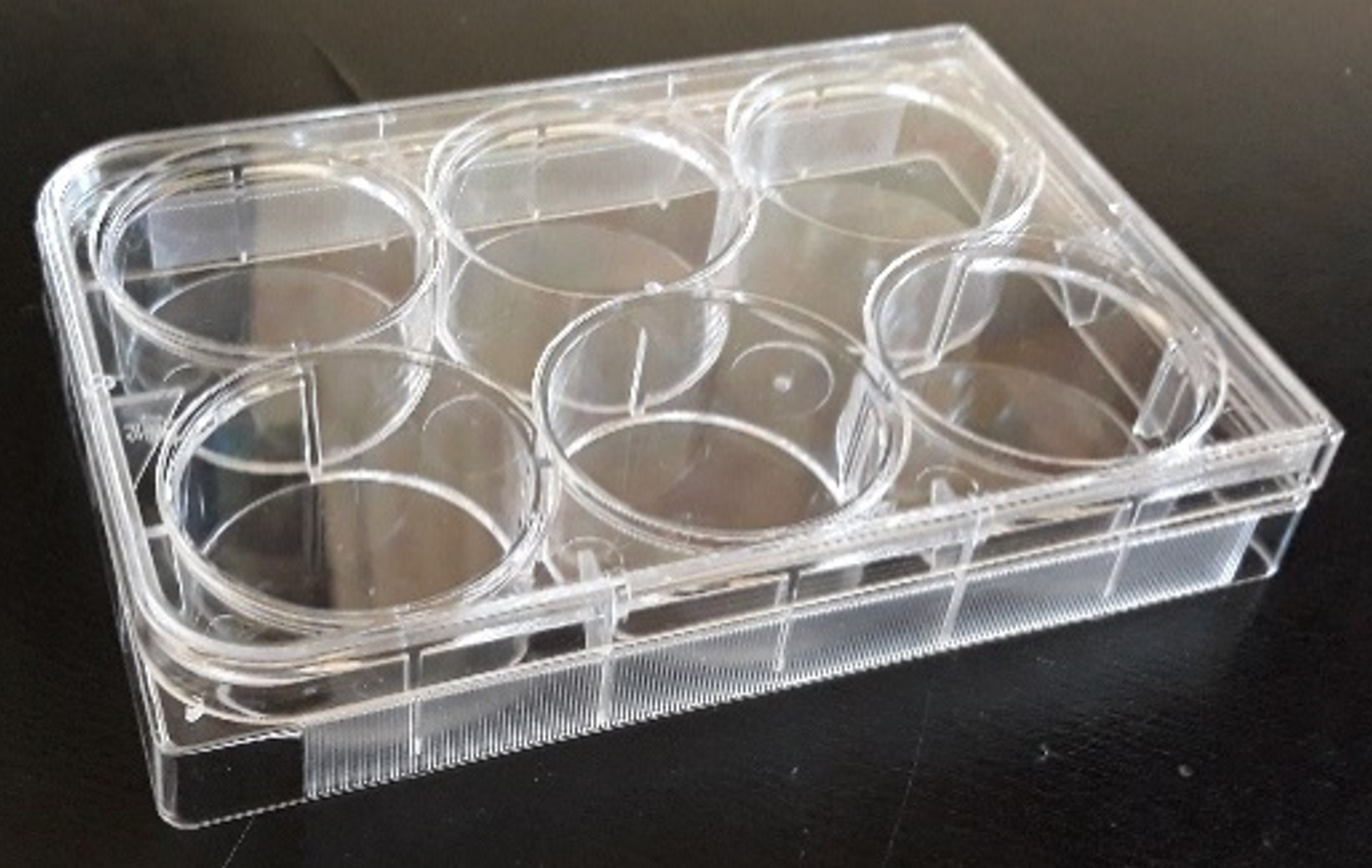}
    \caption{A typical six well cell culture plate.}
    \label{fig:cellSampleContainer}
\end{figure}

The beam is delivered vertically into the end station, which is where the cells will be irradiated. Vertical delivery of the beam allows the use of conventional cell culture plates, such as that shown in Figure~\ref{fig:cellSampleContainer}, that provide breathable wells where the cell sample is grown on the bottom of a well filled with cell nutrient liquid. 
Such cell culture plates are available in a variety of sizes and the final choice will be dependent on the numbers of cell samples required for a particular cell type and the configuration of the beam will be matched such that the whole area of the well will be irradiated.

\begin{figure}[ht]
    \centering
    \includegraphics[width=0.4\linewidth]{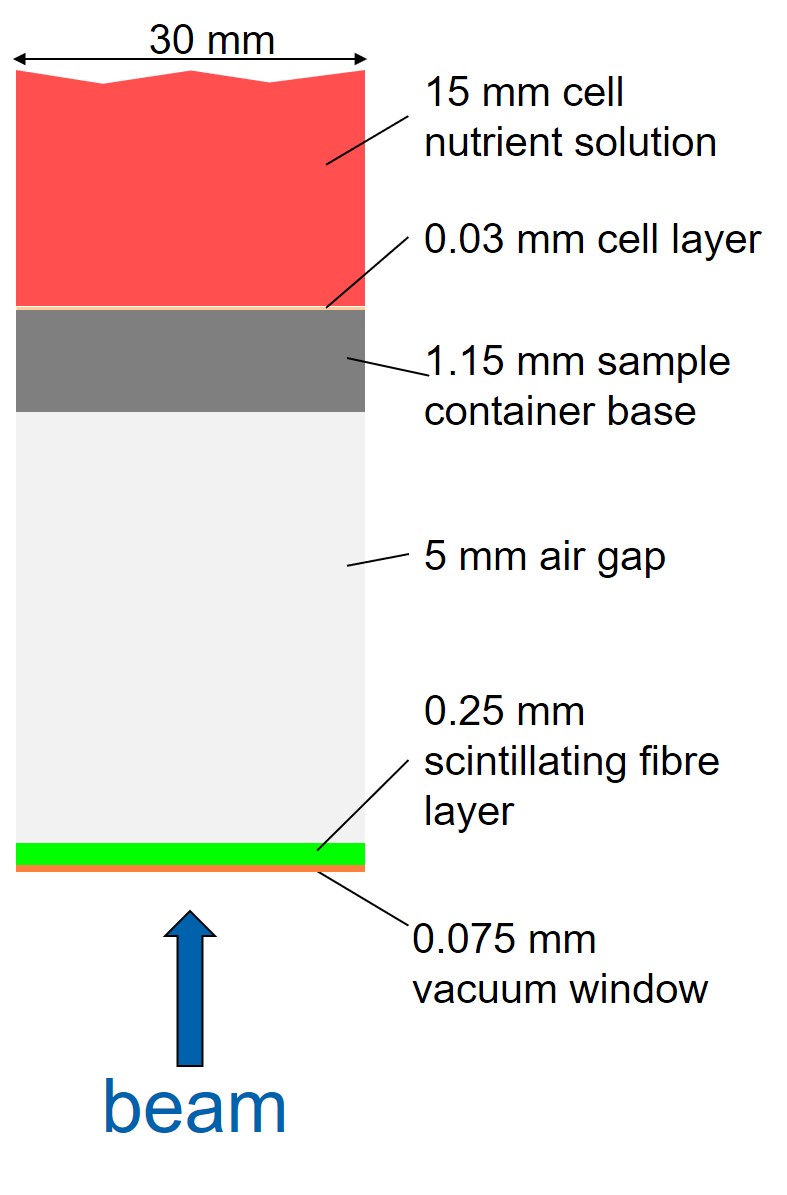}
    \caption{Design of the end station simulated with BDSIM.}
    \label{fig:endStationDiagram}
\end{figure}

Since the beam energy is low, any unnecessary material in the beam path may prevent the beam reaching the cell layer.  Increasing the beam energy is possible but would require a more costly laser system.  Figure~\ref{fig:endStationDiagram} shows the current schematic design of the end station and the materials in the path of the beam. The thickness of the cell culture plate was taken from measurements of the one shown in Figure~\ref{fig:cellSampleContainer}. This could be made thinner by manufacturing custom cell culture plates instead of using off-the-shelf ones, if this was a limiting factor. For the simulation only the area of the nominal beam aperture was considered, i.e. a cylinder 30\,mm in diameter, since the expected beam divergence is very small ($<10^{-4}$\,rad). The scintillating fibre layer will be used to monitor the beam uniformity in real time, pulse-by-pulse. Using two layers would allow determining the transverse dose profile delivered to each cell sample. A prototype detector that can provide a 2D beam intensity profile measurement with a position resolution of 250\,$\mu$m is currently under development.  At high beam intensities the life-time of a fibre based detector could be an issue and this will be investigated by testing the prototype with a laser-driven ion beam. The possibility of using multiple layers (in an offline measurement without the cell culture plate) to accurately predict the dose deposition profile in the cell layer is also being investigated. 

For absolute dose calibration, conventional diagnostics used for laser-driven ion beams, such as radiochromic films (e.g. Gafchromic film) and nuclear track detectors (e.g. CR-39), could be used. By placing the detector layer (i.e. either the Gafchromic film or the CR-39 layer) at the position of the cell layer with material in front identical to the base of the cell culture plate, the dose that would be delivered to the cell layer can be measured. The issue with using such detectors is that these measurements would be done in a separate run to the cell irradiations, meaning variations in beam conditions would affect the actual dose delivered to the cells. However, it may be possible to correct for such variations by correlating the intensity profile measurements from the scintillating fibre layer with the absolute dose measurements.  Calibration of the films is another important issue and would require tests at a facility where the delivered dose can be accurately determined. Other groups have performed similar experiments, see for example~\cite{Reinhardt2015} and~\cite{Bin2019}.

\section{Results}
\begin{figure}[ht]
    \centering
    \includegraphics[width=0.9\linewidth]{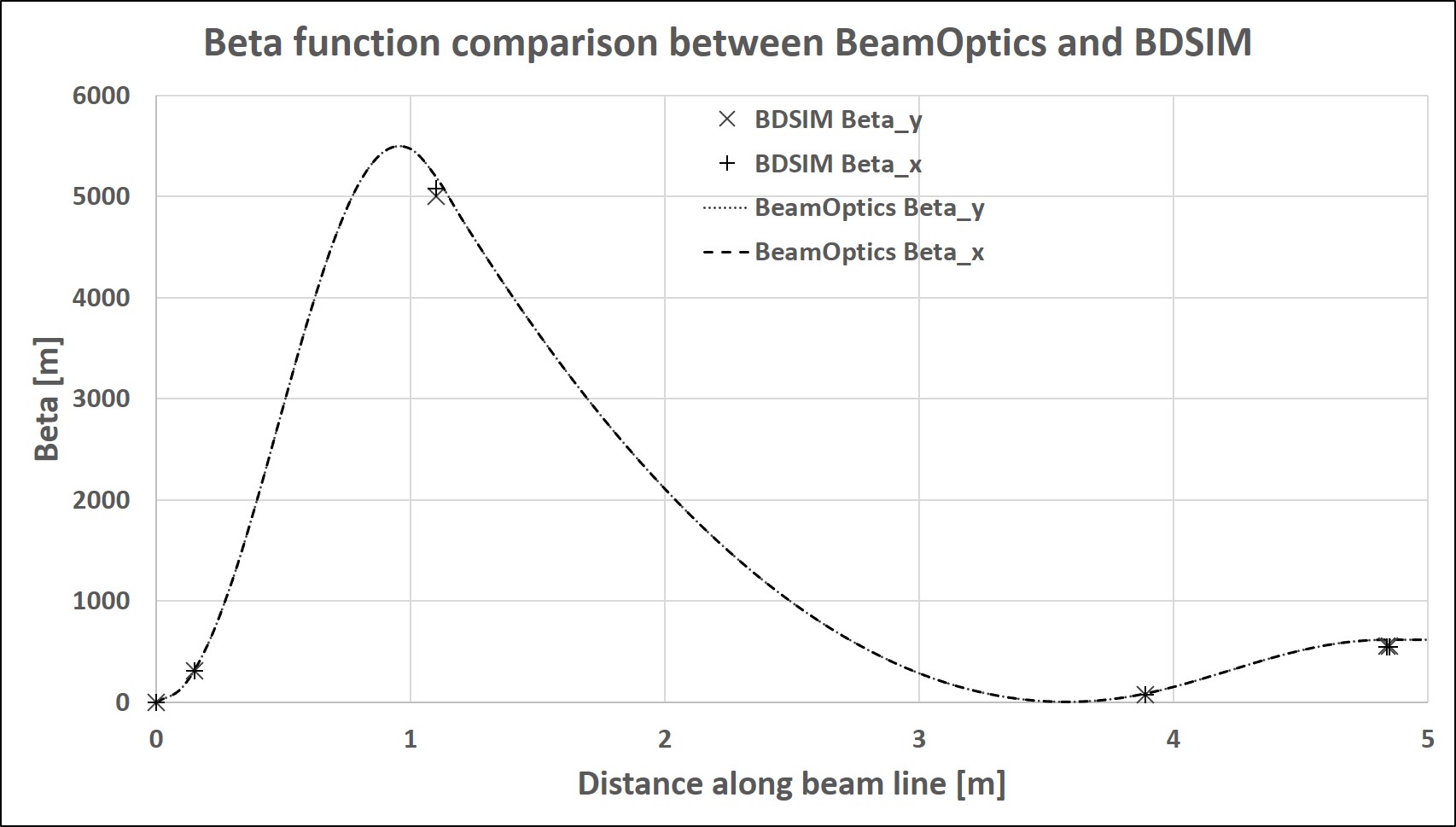}
    \caption{Comparison of the beta function from the beam optics design code BeamOptics and the beta values extracted from the BDSIM simulation for the capture section.}
    \label{fig:captureOptics}
\end{figure}

\begin{figure}[ht]
    \centering
    \includegraphics[width=0.9\linewidth]{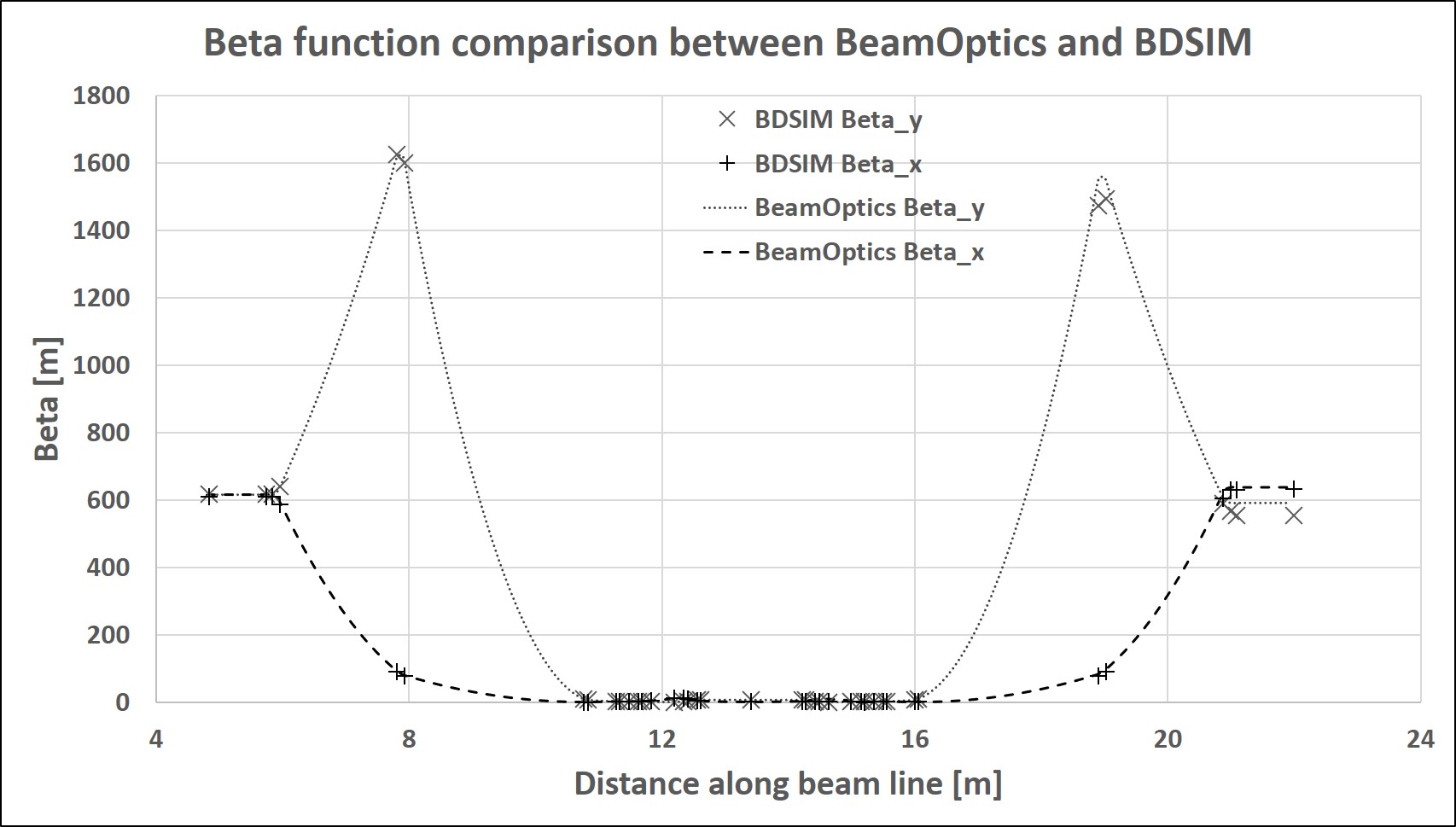}
    \caption{Comparison of the beta functions from BeamOptics and BDSIM for the beam transport section.}
    \label{fig:transportOptics}
\end{figure}

Figures~\ref{fig:captureOptics}~and~\ref{fig:transportOptics} show a comparison of the beta function (which gives the beam envelope as a function of distance along the beam-line) from the beam optics design, which was performed using BeamOptics, and the beta values after each element in the BDSIM simulation.  Figure~\ref{fig:captureOptics} gives the comparison for the capture section and shows good agreement between the calculated beta from BeamOptics and the beta calculated from the particle distributions from the BDSIM tracking simulations.  The Gabor lens was implemented in BDSIM using a solenoid of equivalent focusing strength.  In the future, the Gabor lens will be simulated to produce a 3-D electro-magnetic field map that can then be used for tracking in BDSIM.

Figure~\ref{fig:transportOptics} shows the comparison of the beta function from BeamOptics and BDSIM for the beam transport section.  The beta functions show good agreement and at the end of the transport section the beta function is flat, which means the divergence of the beam in the end station will be small. 

\begin{figure}[ht]
    \centering
    \includegraphics[width=0.9\linewidth]{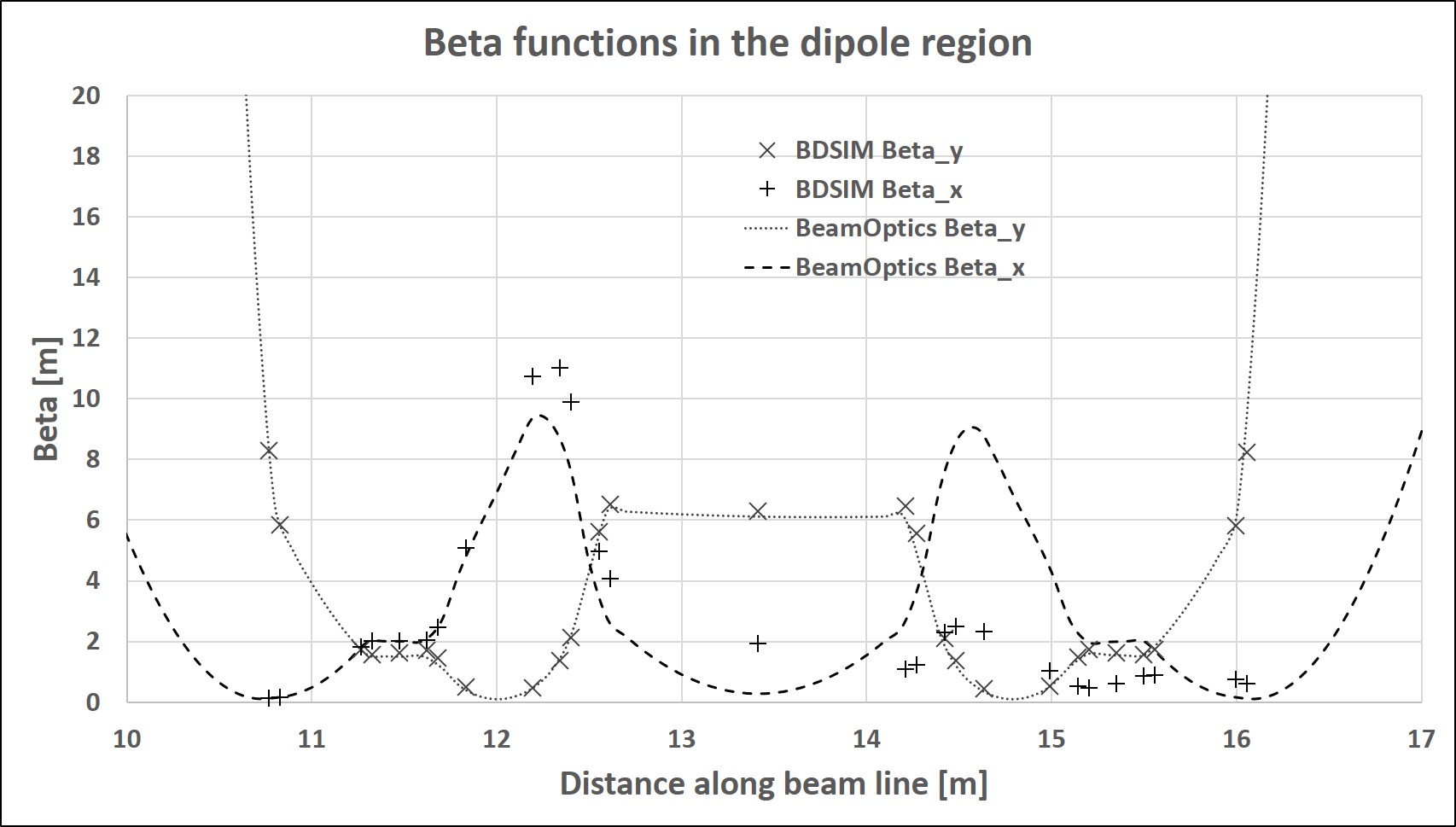}
    \caption{Comparison of the beta functions in the 90$^{\circ}$ bend region.}
    \label{fig:transportOpticsZoom}
\end{figure}

Figure~\ref{fig:transportOpticsZoom} zooms in on the beta functions in the bending region.  There can be seen a mismatch in the beta functions in the x-direction. This is probably due to differences in the treatment of edge focusing in BeamOptics and BDSIM and is currently under investigation, though this does not affect the beam significantly since the beta functions are in good agreement downstream with the mismatch in the beta function at the end of the transport section being 7\% in the x-direction and 1\% in the y-direction. This translates to a difference in the beam envelope of 3\% in the x-direction and less than 0.6\% in the y-direction.

\begin{figure}[ht]
\centering\includegraphics[width=0.8\linewidth]{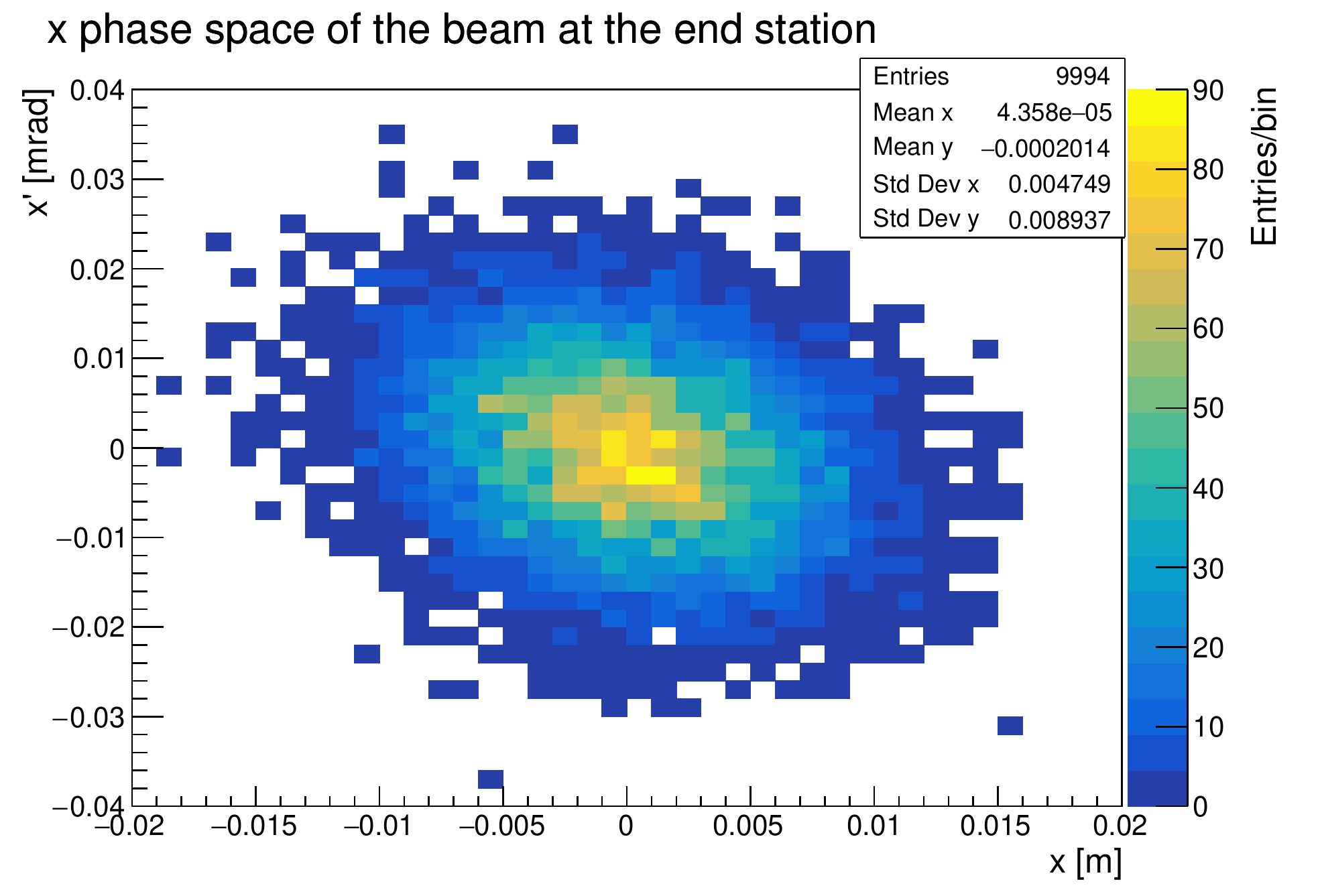}
\caption{Distribution of the x phase space of the beam at the entrance of the end station.}
\label{fig:outXPhaseHist}
\end{figure}

\begin{figure}[ht]
\centering\includegraphics[width=0.8\linewidth]{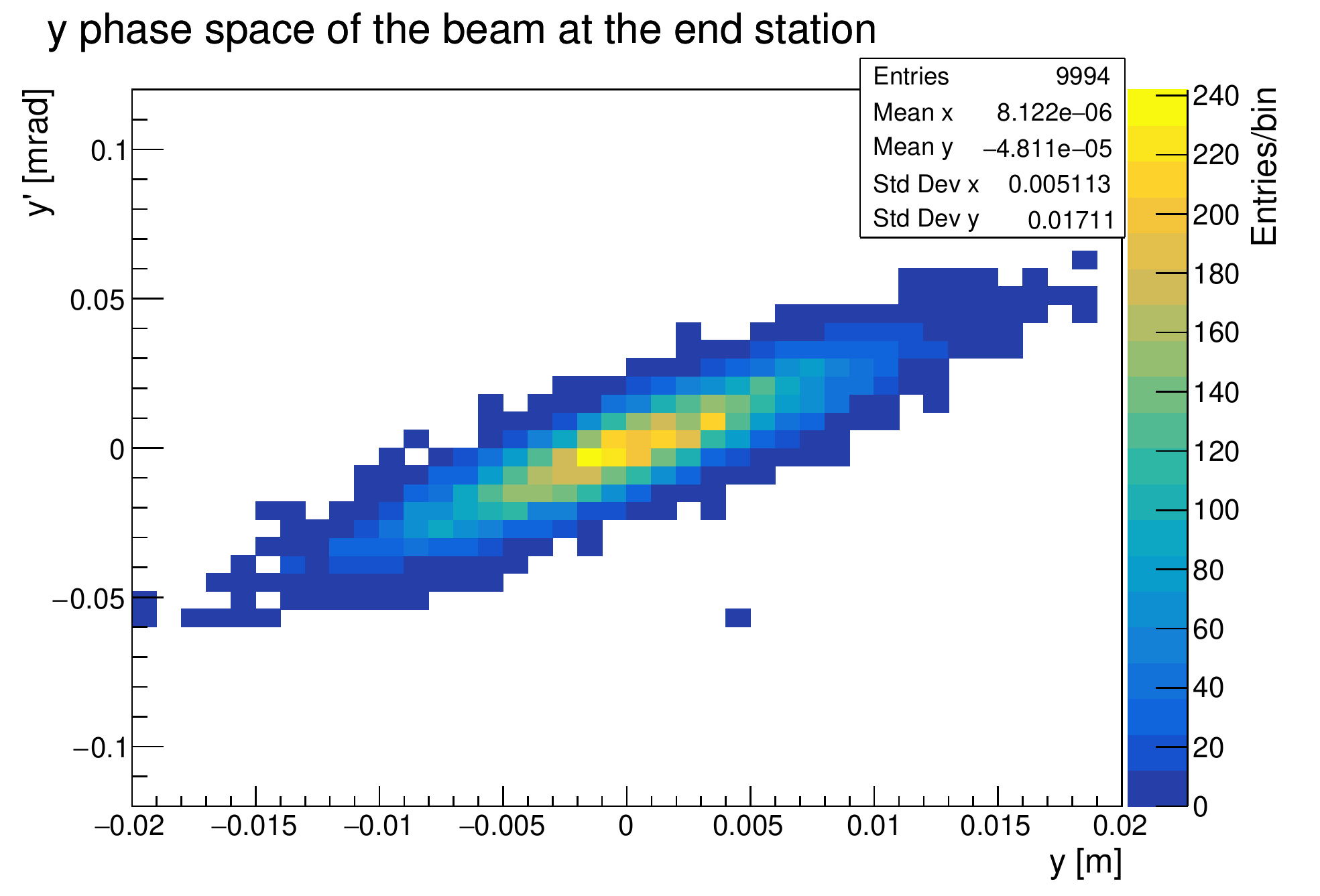}
\caption{Distribution of the y phase space of the beam at the entrance of the end station.}
\label{fig:outYPhaseHist}
\end{figure}

\begin{figure}[ht]
\centering\includegraphics[width=0.8\linewidth]{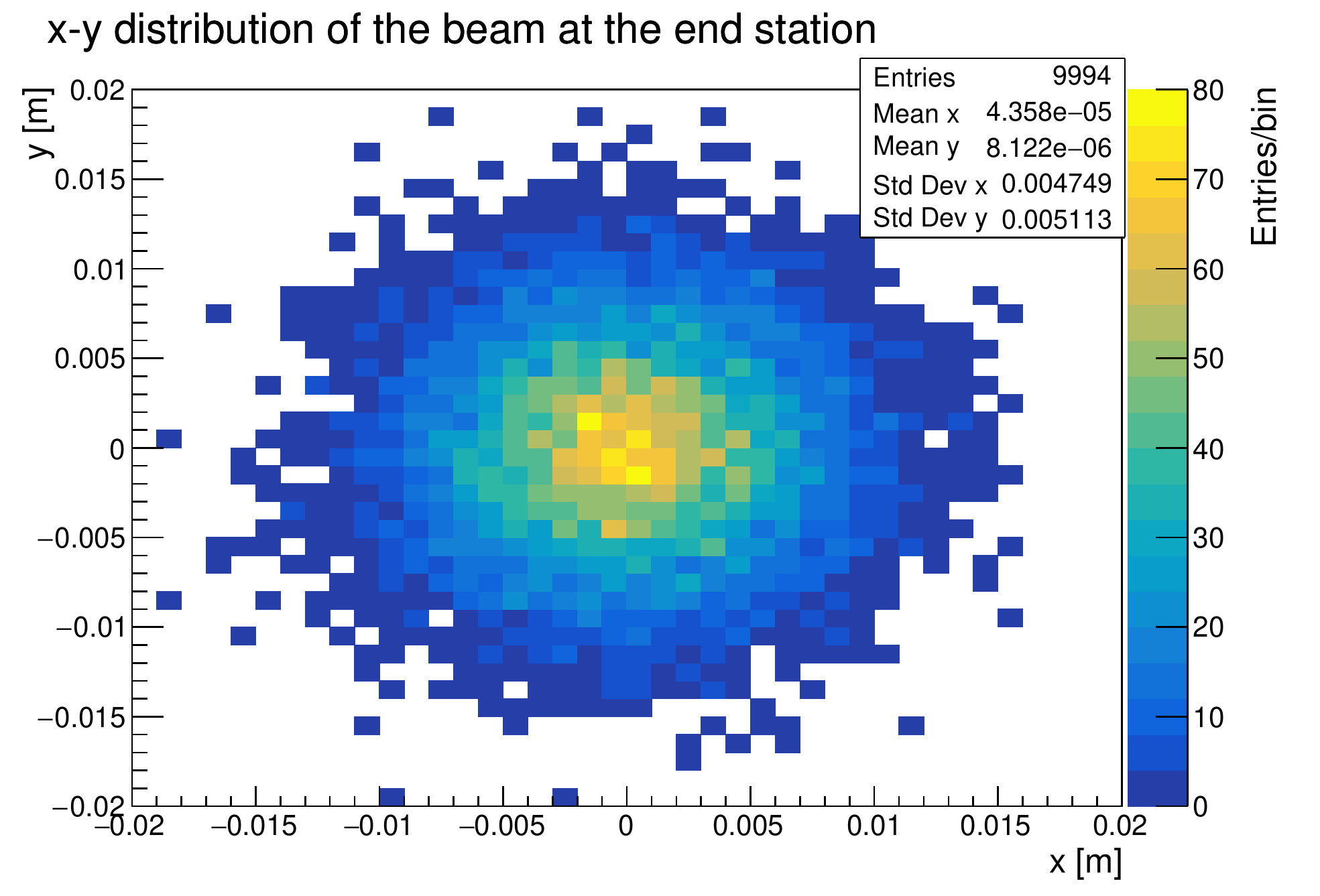}
\caption{Distribution of the x-y profile of the beam at the entrance of the end station.}
\label{fig:outXYHist}
\end{figure}

The distribution of the beam at the entrance of the end station, extracted from the BDSIM simulation, is shown in Figures~\ref{fig:outXPhaseHist},~\ref{fig:outYPhaseHist}~and~\ref{fig:outXYHist}.  The important result here is that the divergence of the beam in the end station is very small, i.e. $<10^{-4}$ and that the transmission is 99.9\% for a simulation run of $10{^4}$ particles.  The x-y profile of the beam is similar to that of the input beam and will determine the dose distribution over the cell sample.  Work is ongoing to investigate the effect of the non-uniform beams produced by the laser on the beam distribution at the end station.  However, the aim for LhARA is to measure this pulse-by-pulse using the scintillating fibre detector so that the actual transverse dose profile delivered to the cells can be determined.

\begin{figure}[ht]
    \centering
    \includegraphics[width=0.9\linewidth]{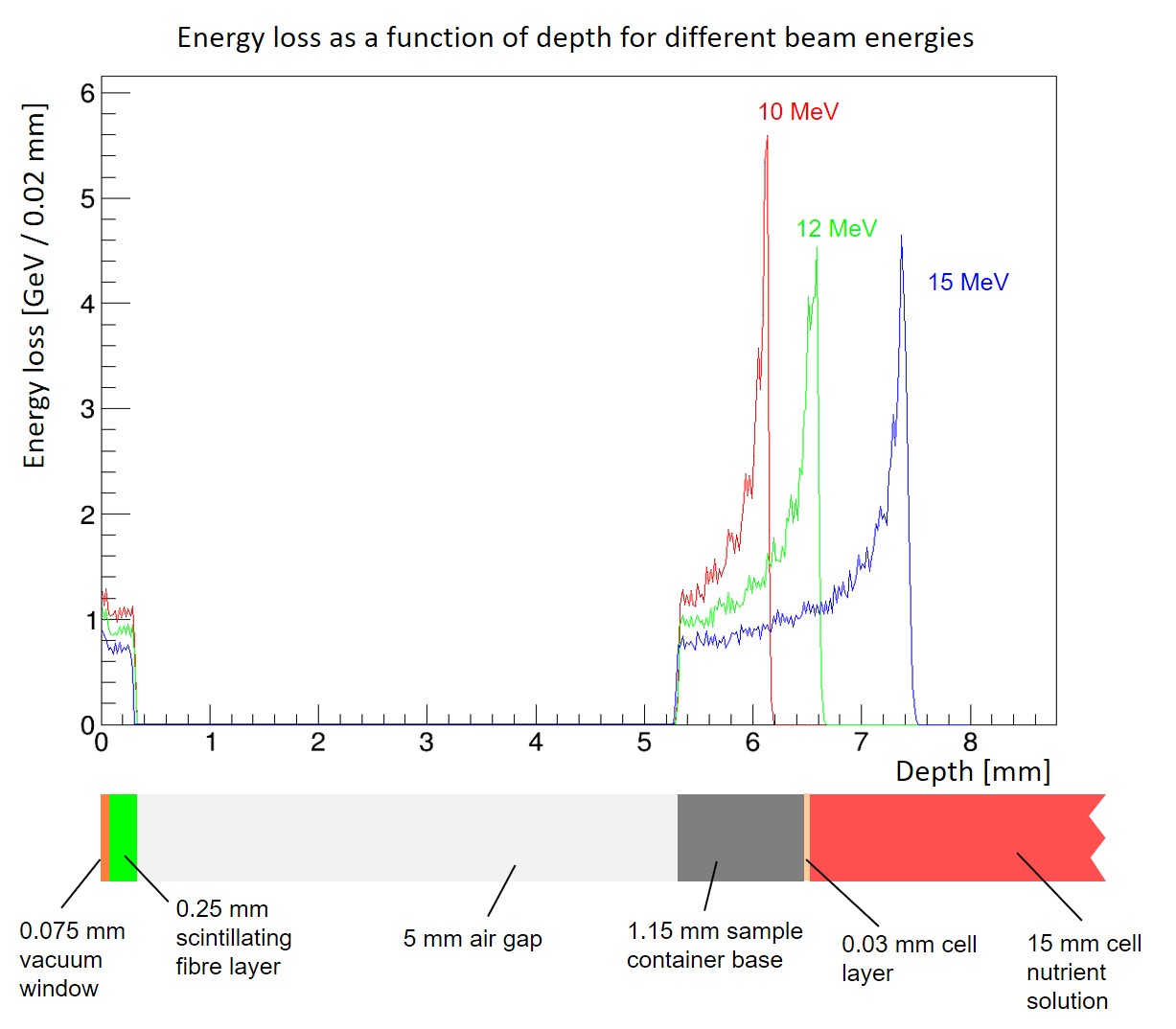}
    \caption{Energy loss as a function of depth in the end station for three different beam energies 10\,MeV, 12\,MeV and 15\,MeV. For each beam energy $1\times10^4$ particles were simulated. The strip below the graph shows the material composition of the end station to scale.}
    \label{fig:energyLoss}
\end{figure}

\begin{figure}[ht]
    \centering
    \includegraphics[width=0.9\linewidth]{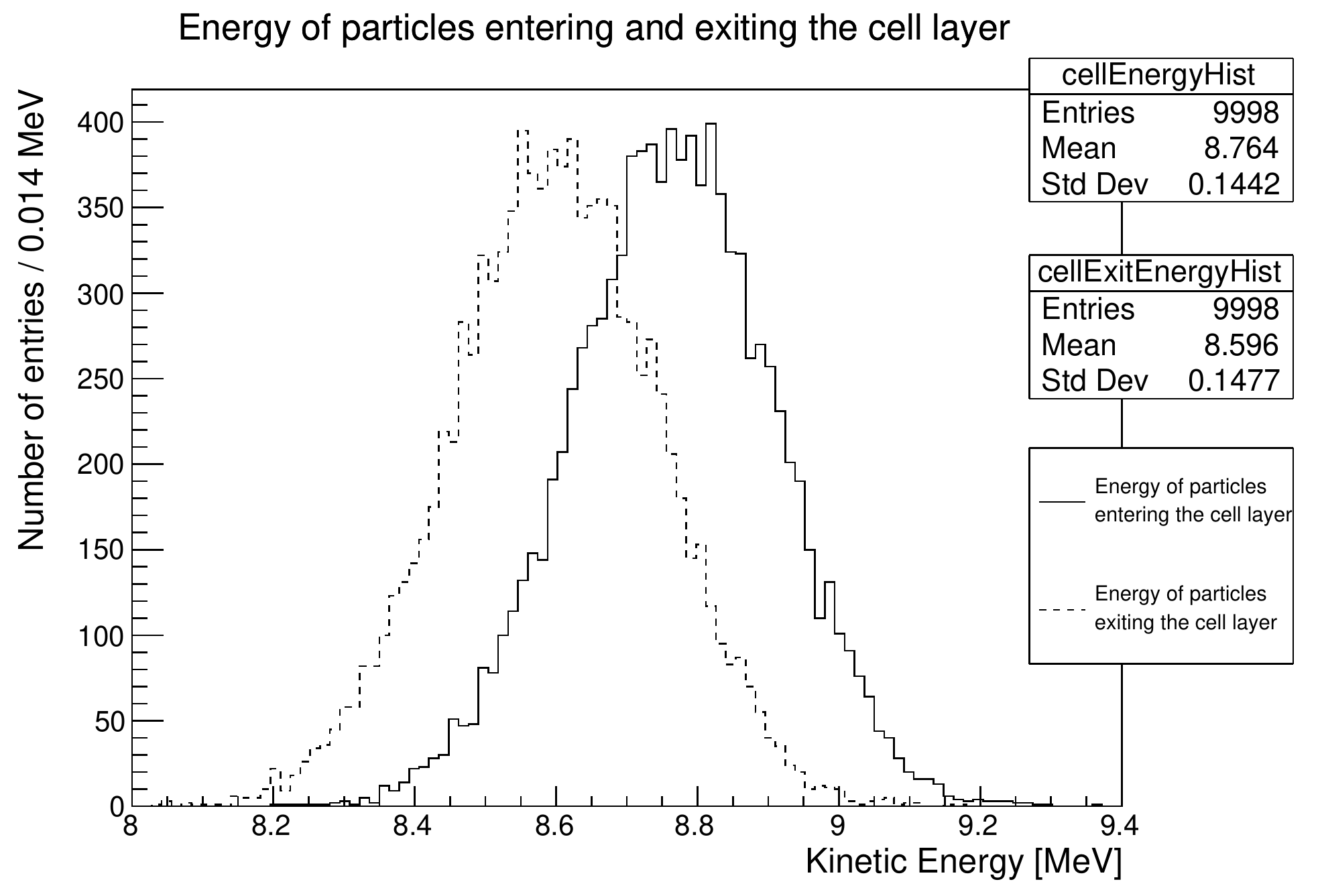}
    \caption{Energy of the particles entering and exiting the cell layer for a 15\,MeV proton beam.}
    \label{fig:cellEnergyHists}
\end{figure}

\begin{figure}[ht]
    \centering
    \includegraphics[width=0.9\linewidth]{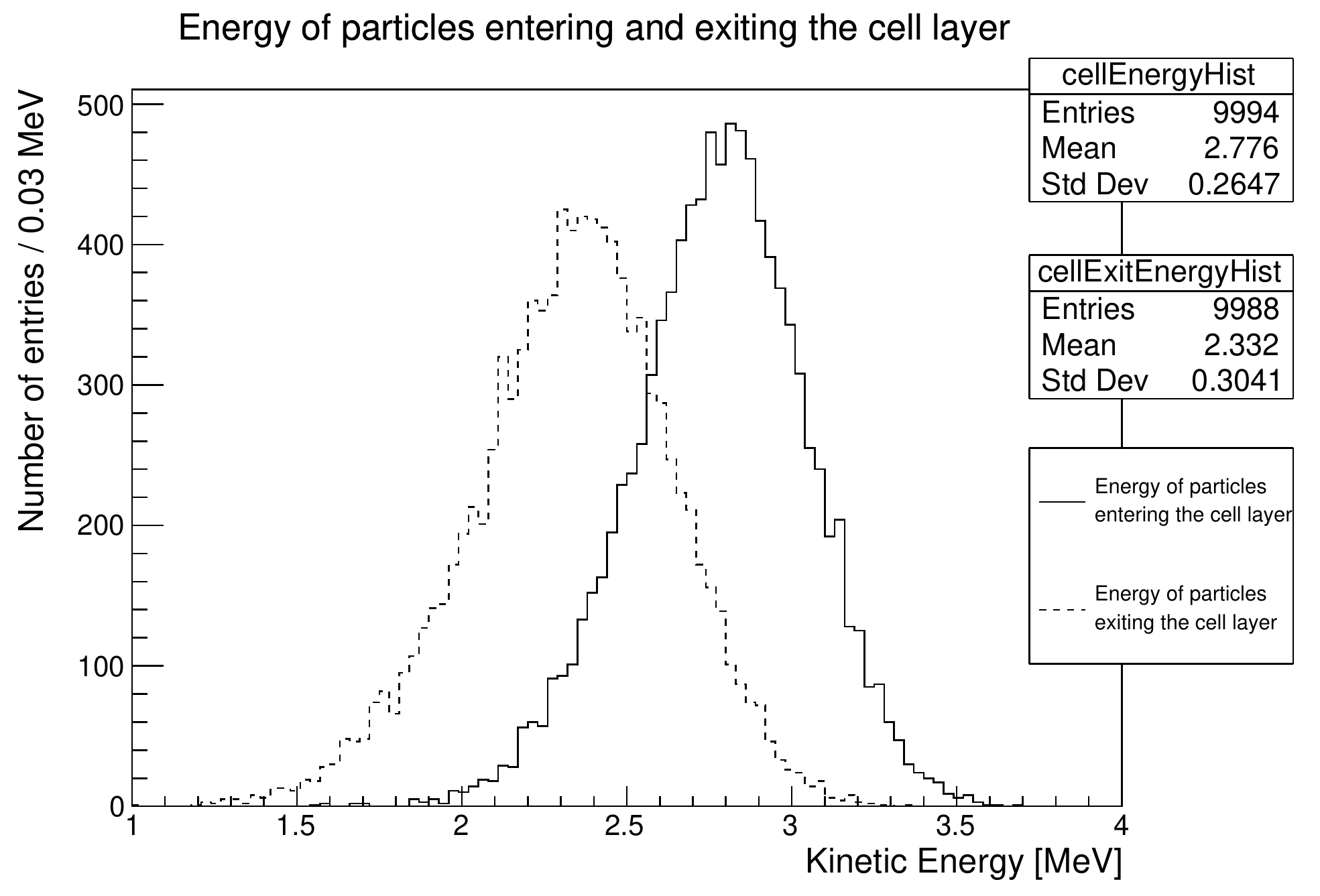}
    \caption{Energy of the particles entering and exiting the cell layer for a 12\,MeV proton beam.}
    \label{fig:cellEnergyHists_12MeV}
\end{figure}

BDSIM was used to extract the energy deposited in the different material layers, which is shown in Figure~\ref{fig:energyLoss} for three different beam energies.  This shows that a beam with kinetic energy of 10\,MeV does not reach the cell layer but the 12\,MeV beam has the Bragg peak (i.e. where most of the energy is deposited) at the position of the cell layer. However, this would not allow investigating the radiobiological effect of irradiating a sample located before the Bragg peak.  Thus the design kinetic energy of 15\,MeV has been chosen since it will also contain a significant number of particles in the range 10--15\,MeV allowing irradiation with a variety of energies.
The energies of the particles entering and exiting the cell layer for a 15\,MeV proton beam is given in Figure~\ref{fig:cellEnergyHists}.  This shows the mean energy of particles entering the cell layer is 8.8\,MeV with an energy spread (caused by stochastic effects) of 0.14\,MeV.  From the two plots in Figure~\ref{fig:cellEnergyHists} the energy deposited in the cell layer was calculated to be 1.7\,GeV ($1.2\times10^{-5}$\,Gy) for a bunch of $10^4$ protons.  For the 12\,MeV beam the energies of the particles entering and exiting the cell layer is given in Figure~\ref{fig:cellEnergyHists_12MeV}. In this case the energy deposited in the cell layer is 4.5\,GeV ($3.1\times10^{-5}$\,Gy) for $10^4$ protons, which is much higher than for the 15\,MeV beam, as expected, since the cell layer is located in the Bragg peak for the 12\,MeV beam.  However, the actual number of protons per pulse for LhARA could be significantly higher than this, $10^6$--$10^9$ protons. Thus the dose per pulse could be up to 1.2\,Gy for 15\,MeV protons and 3.1\,Gy for 12\,MeV protons and is comparable to the typical dose per fraction for a clinical treatment but delivered in a single pulse.

\section{Discussion and conclusions}
\label{sec:summary}
The CCAP has proposed a concept for a new facility to perform detailed systematic studies of cell irradiation.  The facility will use novel accelerator technologies to deliver intense beams of protons and ions from helium to carbon.  The optics design of the beam-line for the first stage of LhARA has been verified by comparing the beam envelope calculations from BeamOptics with particle tracking simulations using BDSIM and these show good agreement. The energy deposition profile in the end station has also been investigated and this shows that for the current design, 15\,MeV is a good choice for the laser cut-off energy allowing irradiating the cells before and within the region of the Bragg peak.

In these simulations, space charge effects have not been included since BDSIM does not include this effect as it is based on Geant4 which is a single-particle tracking code.  This is clearly an important issue for the high-intensity bunches produced by the laser system and the capture system.  Verification of these results will be done using a different simulation tool that includes space charge effects.

Further work is on-going to finalise the design of LhARA including: updating the BDSIM simulations with input beams generated directly from EPOCH simulations; improving the description of the Gabor lens in BDSIM using a 3D electro-magnetic field map; investigating a more compact beam transport channel; and designing LhARA Stage\,II that will include a post-accelerator to deliver higher beam energies for in vivo studies with protons and in vitro studies with heavier ions.

\clearpage

\section*{Acknowledgements}
\label{sec:ack}
The authors would like to thank MedAustron and the Medical University of Vienna for their support.

This work was supported by the Engineering and Physical Sciences Research Council [Grant No. EP/K022415/1] and the Science and Technology Facilities Council [Grant Nos. ST/P002021/1 and ST/N000242/1].

The EPOCH code was developed as part of the UK EPSRC funded projects EP/G054940/1.





\section*{References}
\bibliographystyle{model1-num-names}
\bibliography{bibliography.bib}







\end{document}